\documentclass[conference,compsoc]{IEEEtran}
\usepackage{graphicx}
\usepackage{multirow}

\newcommand{\green}[1]{\textcolor{mygreen}{#1}}
\newcommand{\red}[1]{\textcolor{myred}{#1}}

\usepackage{amsmath}
\usepackage{bbm}
\usepackage{booktabs}   
\usepackage{multirow}   
\usepackage{cleveref}
\usepackage{placeins}
\usepackage{afterpage}

\usepackage{xcolor}
\definecolor{mygreen}{rgb}{0.0, 0.7, 0.0} 
\definecolor{myred}{rgb}{0.8, 0.0, 0.0}   

\usepackage[justification=justified,singlelinecheck=false]{caption}

\ifCLASSOPTIONcompsoc
  \usepackage[nocompress]{cite}
\else
  \usepackage{cite}
\fi

\ifCLASSINFOpdf
\else
\fi
\begin{document}

\title{vEcho: A Paradigm Shift from Vulnerability Verification to Proactive Discovery with Large Language Models}










\author{
\IEEEauthorblockN{
Mingcheng Jiang\textsuperscript{1*}, 
Jiancheng Huang\textsuperscript{1*}, 
Jiangfei Wang\textsuperscript{2}, 
Zhengzhu Xie\textsuperscript{1}, \\
Nan Fang\textsuperscript{3}, 
Guang Cheng\textsuperscript{1}, 
Xiaoyan Hu\textsuperscript{1}, and 
Hua Wu\textsuperscript{1\textdagger}
}

\IEEEauthorblockA{\textsuperscript{1}\textit{Southeast University}\\
Emails: slainnico@gmail.com, shoucheng@seu.edu.cn, 220255762@seu.edu.cn, \\
chengguang@seu.edu.cn, xyhu@njnet.edu.cn, hwu@seu.edu.cn}

\IEEEauthorblockA{\textsuperscript{2}\textit{Anhui University}\\
Email: lecrepuscular@gmail.com}

\IEEEauthorblockA{\textsuperscript{3}\textit{Anhui Shenwu Information Technology Co., Ltd.}\\
Email: me@yvling.cn}

\thanks{\textsuperscript{*}These authors contributed equally to this work.}
\thanks{\textsuperscript{\textdagger}Corresponding author.}
}


%


\maketitle

\begin{abstract}
Static Application Security Testing (SAST) tools often suffer from high false positive rates, leading to alert fatigue that consumes valuable auditing resources. Recent efforts leveraging Large Language Models (LLMs) as filters offer limited improvements; however, these methods treat LLMs as passive, stateless classifiers, which lack project-wide context and the ability to learn from analyses to discover unknown, similar vulnerabilities.
In this paper, we propose vEcho, a novel framework that transforms the LLM from a passive filter into a virtual security expert capable of learning, memory, and reasoning. vEcho equips its core reasoning engine with a robust developer tool suite for deep, context-aware verification. More importantly, we introduce a novel Echoic Vulnerability Propagation (EVP) mechanism. Driven by a Cognitive Memory Module that simulates human learning, EVP enables vEcho to learn from verified vulnerabilities and proactively infer unknown, analogous flaws, achieving a paradigm shift from passive verification to active discovery.
Extensive experiments on the CWE-Bench-Java dataset demonstrate vEcho's dual advantages over the state-of-the-art baseline, IRIS. Specifically, vEcho achieves a 65\% detection rate, marking a 41.8\% relative improvement over IRIS's 45.83\%. Crucially, it simultaneously addresses alert fatigue by reducing the false positive rate to 59.78\%, a 28.3\% relative reduction from IRIS's 84.82\%. Furthermore, vEcho proactively identified 37 additional known vulnerabilities beyond the 120 documented in the dataset, and has discovered 51 novel 0-day vulnerabilities in open-source projects.
\end{abstract}


%
\IEEEpeerreviewmaketitle

\section{Introduction}

Static Application Security Testing (SAST) is an indispensable component of the modern Software Development Life Cycle (SDLC). However, SAST tools have long been notoriously plagued by high false positive rates, leading to severe Alert Fatigue \cite{johnson2013don}\cite{pearce2023examining}\cite{ullah2024llms}. This phenomenon critically consumes valuable manual security auditing resources and may cause genuine threats to be overlooked amidst the noise.

To mitigate this issue, recent efforts have emerged that leverage Large Language Models (LLMs) to assist in vulnerability analysis\cite{li2024iris}\cite{sun2024llm4vuln}\cite{deng2024pentestgpt}. The contributions of works represented by IRIS are primarily twofold: 1) attempting to utilize LLMs to infer new taint specifications, thereby expanding the detection scope of SAST; and 2) employing LLMs as filters to verify vulnerability alerts, which addresses alert fatigue to a certain extent \cite{li2024enhancing}.

However, these methods suffer from fundamental methodological limitations. We contend that while they optimize SAST, they fail to transcend the SAST paradigm \cite{bohme2025software}. They treat LLMs as passive, stateless classifiers, which leads to three core deficiencies:

\begin{itemize}
    \item \textbf{Passive, Alert-Driven Mechanism.} In existing methods (e.g., IRIS), the LLM is not an active auditor but rather a passive consumer of SAST tools (e.g., CodeQL \cite{avgustinov2016ql}). Its analytical horizon is entirely constrained by the potentially incomplete or partial taint paths pre-supplied by CodeQL. This fundamentally limits its capacity for exploratory auditing \cite{luo2022tchecker}\cite{wang2023conftainter}.
    
    \item \textbf{Stateless Analysis Process.} Current methods are one-shot and stateless. Although rules are dynamically generated for each project, this generation stems merely from the LLM's own pre-trained knowledge, rather than from an understanding derived from the project itself \cite{ullah2024llms}. They cannot accumulate experience from successful or failed verifications, nor can they transfer patterns learned in Project A to Project B \cite{qian2024chatdev}.
    
    \item \textbf{Capability Boundary Limited to Verification.} The existing workflow terminates immediately after verifying an alert\cite{kim2022learning}. These methods fail to leverage the rich information contained within true positives (TPs) or even false positives (FPs) (e.g., specific developer coding habits or the presence of sanitizers)\cite{chen2021boosting}. Consequently, their capability boundary is locked at verification and cannot extend to proactive discovery.
\end{itemize}

To truly unlock the potential of LLMs, we advocate for a paradigm shift in the way they are utilized. We should not treat the LLM as a passive SAST optimizer but rather construct it as an active virtual security expert—an agent capable of simulating human expert learning, memory, and reasoning, thereby significantly alleviating the workload of human experts and enhancing audit efficiency.

To this end, we propose vEcho, a novel framework designed to realize this vision. vEcho addresses the aforementioned limitations through two core innovations. First, it is equipped with a robust developer tool suite (including code navigation, project comprehension, and web search). When performing deep, context-aware verification, vEcho not only invokes these tools but also actively queries its Cognitive Memory Module to leverage historical audit experience and current project-specific context, thereby achieving high-precision judgments.

More importantly, we introduce a novel Echoic Vulnerability Propagation (EVP) mechanism. This mechanism, also driven by the Cognitive Memory Module, enables vEcho to learn from verified vulnerabilities (TPs) and false positives (FPs) and to infer other unknown, analogous flaws within the codebase proactively. This achieves the paradigm shift from passive verification to active discovery.

The main contributions of this paper are as follows:

\begin{enumerate}
    \item \textbf{A Framework for Deep Context-Aware Vulnerability Auditing.} We propose vEcho, an agent framework that enables an LLM to transcend partial taint paths, actively acquiring and integrating project-level context (such as macro-architecture and security configurations) to achieve high-precision vulnerability verification.
    
    \item \textbf{An Experience-Driven Mechanism for Proactive Discovery:} We design and implement the Echoic Vulnerability Propagation (EVP) mechanism, augmented by the Cognitive Memory Module. This mechanism allows vEcho to learn from verified true positives (TPs) and false positives (FPs), generate new audit directives (i.e., Re-Scan Guidance), and proactively discover unknown, analogous vulnerabilities, achieving a capability breakthrough from passive verification to active discovery.
    
    \item \textbf{Comprehensive Evaluation and SOTA Superiority:} We conducted a rigorous evaluation on the CWE-Bench-Java dataset. The experiments demonstrate vEcho's dual advantages over the SOTA baseline, IRIS: 1) \textit{Improved Detection Rate:} vEcho achieves a 65\% detection rate, marking a 41.8\% relative improvement over IRIS's 45.83\%. 2) \textit{Mitigation of Alert Fatigue:} vEcho reduces the false positive rate (FPR) to 59.78\%, a 28.3\% relative reduction from IRIS's 84.82\%.
    
    \item \textbf{Real-World Impact:} vEcho has achieved significant vulnerability discovery results in practical applications. Not only did vEcho identify 37 additional known vulnerabilities in the CWE-Bench-Java dataset that were not marked by the benchmark, but it also successfully discovered 51 novel 0-day vulnerabilities during audits of other open-source projects. Critically, these discoveries include high-impact Critical Remote Code Execution (RCE) and Arbitrary Code Execution (ACE) vulnerabilities, which have been officially confirmed by the Apache project, fully demonstrating vEcho's exceptional capability in discovering high-impact security flaws.
\end{enumerate}

\section{Background \& Related Work}
This section outlines the fundamental limitations of traditional SAST, explores the emerging applications of LLMs in security tasks, and precisely situates vEcho's unique contributions relative to state-of-the-art (SOTA) LLM-assisted static analysis efforts.

\subsection{Limitations of Traditional Static Application Security Testing (SAST)}
Static Application Security Testing (SAST) is a white-box security testing methodology that discovers security vulnerabilities by analyzing an application's source code or binaries. Mainstream SAST tools, whether open-source (e.g., CodeQL\cite{avgustinov2016ql}) or commercial (e.g., Coverity\cite{synopsys2025coverity}), fundamentally rely on predefined rules\cite{yamaguchi2014modeling}, data-flow analysis, and taint-tracking techniques\cite{livshits2005finding}\cite{tripp2009taj}\cite{arzt2014flowdroid}.

Despite their widespread adoption, these tools suffer from two long-standing, fundamental challenges:
\begin{itemize}
    \item \textbf{High False Positives:} SAST tools often lack a deep understanding of the project's holistic business logic, framework configurations, and custom sanitation functions. This causes them to flag numerous code paths that pose no threat in the actual context, thereby generating severe Alert Fatigue \cite{yang2019towards}.
    \item \textbf{High False Negatives:} Traditional SAST heavily depends on manually crafted and maintained vulnerability rulebases. However, in modern complex software systems (especially within the Java ecosystem), attempting to manually maintain a comprehensive and precise set of taint specifications is an exceedingly arduous and error-prone task\cite{christodorescu2007mining}\cite{chibotaru2019scalable}. When projects adopt new frameworks or novel vulnerability patterns emerge, the lag in the rulebase inevitably leads to missed vulnerabilities.
\end{itemize}

vEcho is designed precisely to address these two core challenges. To combat Alert Fatigue from high false positives, vEcho provides deep verification capabilities based on project-level contextual understanding, identifying genuine vulnerabilities and filtering false positives. To combat high false negatives, vEcho introduces the EVP mechanism to proactively discover unknown, analogous vulnerabilities caused by similar coding patterns that exist outside the SAST rulebase.

\subsection{Emerging Applications of LLMs in Security Tasks}
The powerful capabilities demonstrated by LLMs in code comprehension and reasoning have rapidly made them a research hotspot in the security domain.

\subsubsection*{LLMs as Security Testing Agents} Recent research has begun to explore LLMs as the core drivers of automated security testing. For example, PentestGPT \cite{deng2024pentestgpt} constructs an LLM as a penetration testing agent; JailFuzzer \cite{dong2025fuzz} utilizes LLM-based Agents to automatically generate fuzzing prompts; and YuraScanner \cite{stafeev2024yurascanner} employs LLMs for task-driven dynamic black-box scanning (DAST). These works compellingly demonstrate the significant potential of LLMs as agents in dynamic testing and black-box penetration. vEcho borrows this LLM-centric agent philosophy but innovatively applies it to the static analysis (SAST) domain, creating a white-box audit agent.

\subsubsection*{Evaluation of LLM's Code Security Reasoning} Another line of research evaluates the inherent code security understanding capabilities of LLMs. For instance, research on SV-TrustEval-C \cite{li2025sv} indicates that current LLMs rely more on pattern matching than genuine logical reasoning in vulnerability analysis. Concurrently, other studies have demonstrated that providing LLMs with richer context (e.g., data-flow and control-flow) significantly enhances their detection performance \cite{yang2025context}\cite{zhou2019devign}\cite{lu2024grace}. These evaluations provide critical justification for vEcho's design. The design philosophy of vEcho is based on precisely this insight: to obtain reliable analytical results, the LLM's reasoning process must be anchored in a robust developer tool suite and structured context (such as a knowledge base).

\subsection{LLM-Assisted Static Analysis}
The work most directly related to vEcho combines LLMs with SAST.

\subsubsection*{Domain-Specific LLM+SAST} Existing works have successfully applied this combined approach to specific domains, such as GPTScan \cite{sun2024gptscan} for detecting smart contract vulnerabilities and LLIFT \cite{li2024enhancing} for analyzing Linux kernel code. These works validate the efficacy of the LLM+SAST approach for specific languages and vulnerability types, but their core methodologies are tightly coupled to their specific target domains. In exploring a general paradigm, KNighter \cite{yang2025knighter} proposed a novel paradigm: utilizing LLMs to learn from historical patches and automatically synthesize new static analysis checkers. vEcho's objective differs; we aim to build an extensible audit framework that simulates a human expert's code audit process to uncover unknown vulnerabilities.

\subsubsection*{The SOTA Baseline} The most relevant work to ours is IRIS \cite{li2024iris}, which systematically combines LLMs with CodeQL to detect Java Web vulnerabilities, aligning with our target domain. IRIS's core contribution lies in using LLMs to automate the generation of taint analysis specifications (sources and sinks) required by CodeQL. Its workflow is as follows: first, the LLM analyzes APIs to infer rules; second, CodeQL scans using these rules; finally, the LLM intervenes again to analyze the paths reported by CodeQL to verify if they are false positives.

Despite IRIS being a significant advancement, it remains trapped in the old paradigm of optimizing SAST \cite{bessey2010few} and suffers from the three core limitations we identified in the Introduction:
\begin{itemize}
    \item \textbf{Paradigm Limitation:} IRIS is, in essence, a smarter SAST rule engine. Its entire workflow remains passive and alert-driven. Its context is confined to the isolated taint-path provided by CodeQL \cite{codeql_path_queries}. In contrast, vEcho is LLM-agent-centric, where SAST serves as merely one tool for providing initial leads. 
    vEcho can actively invoke its \textbf{Developer Tool Suite} to acquire project-level context—such as understanding the project's framework, version, or even the specific functionality of certain functions. This is a capability IRIS is architecturally incapable of achieving.
    \item \textbf{Lack of Learning and Memory:} IRIS's analysis is memoryless and one-shot. It cannot transfer insights gained in one project (e.g., the hazardous usage of a specific library) to another. In contrast, vEcho's Cognitive Memory Module enables cross-project knowledge accumulation and transfer.
    
    \item \textbf{One-Shot Task:} IRIS's capability boundary is verification of CodeQL-scanned paths; its process terminates after verification. It cannot learn from confirmed vulnerabilities (or even from false positives) to proactively discover other, logically analogous vulnerabilities in the codebase that were not scanned. vEcho's EVP mechanism is designed precisely to fill this fundamental gap.
\end{itemize}

In summary, while work like IRIS has pioneered the LLM+SAST field, it remains constrained by a passive, stateless paradigm\cite{yu2025stateful}\cite{agrawal2025llm}. vEcho is not an incremental improvement on these methods but a fundamental methodological innovation.

  \section{Design of vEcho}
This section elaborates on the architectural design of vEcho. We first introduce the core insights driving vEcho's design and its overall framework in Section 3.1. Subsequently, we will detail the framework's four core stages: Vulnerability Candidate Generation and Filtering (Section 3.2), Context-Aware Deep Verification (Section 3.3), Cognitive Learning and Feedback (Section 3.4), and the core innovation, the Echoic Vulnerability Propagation (EVP) mechanism (Section 3.5).

\begin{figure*}
    \centering
    \includegraphics[width=1\textwidth]{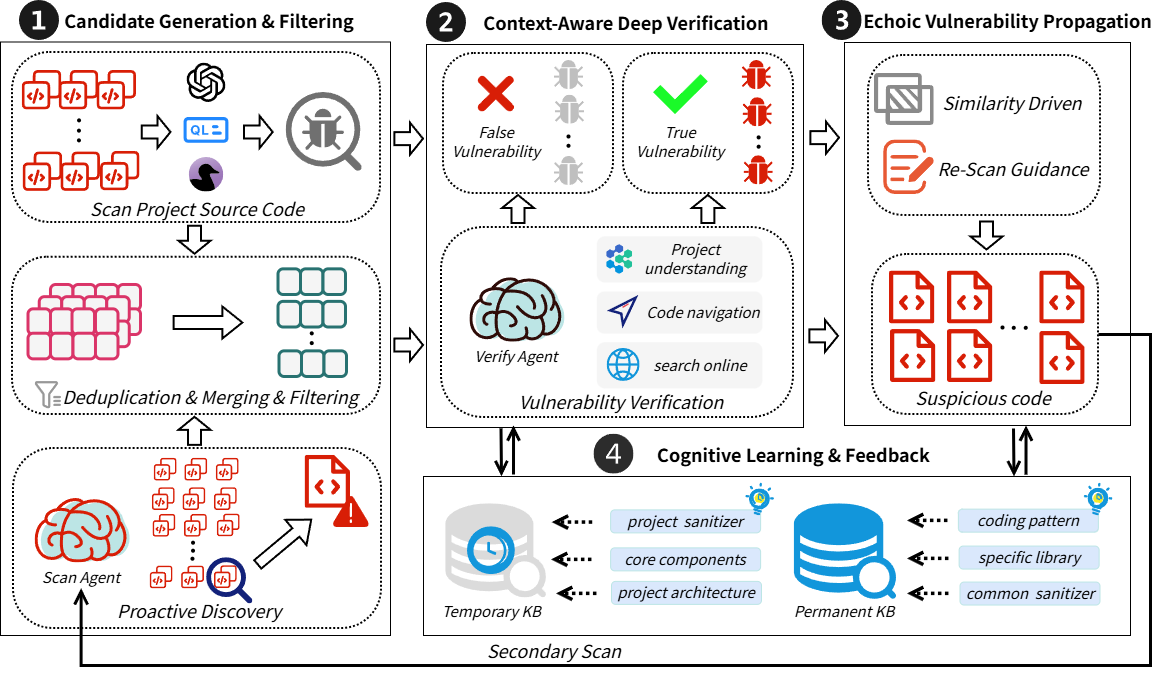}
    \caption{Overview of the vEcho framework. 
    The architecture illustrates an intelligent closed loop comprising four stages: (1) Candidate Generation \& Filtering, (2) Context-Aware Deep Verification, and (3) Echoic Vulnerability Propagation. This entire process is orchestrated and enhanced by the (4) Cognitive Learning \& Feedback module, which feeds learned patterns and new suspicious code back into the verification queue.}
    \label{fig:framework}
\end{figure*}

\subsection{Key Observation}
The design of vEcho is predicated on two core insights, aimed at overcoming the fundamental limitations of existing LLM-assisted analysis tools.

\subsubsection*{Insight 1: Precise vulnerability verification relies on a deep investigation of project-level context} The limitation of SAST alerts lies in their lack of context \cite{zhong2023scalable}. We observe that senior security experts, when verifying a vulnerability, actively trace data flows, inspect global project configurations, and incorporate the latest knowledge from the web. This process demonstrates that simple alert judgment is insufficient. A high-precision LLM code audit framework must transition from a passive arbiter to an active investigator, possessing the ability to access and analyze multi-dimensional project context, rather than merely acting as a passive code snippet analyzer.
    
\subsubsection*{Insight 2: Verified alerts (both TPs and FPs) are invaluable fingerprints for guiding proactive vulnerability discovery} Vulnerabilities often follow specific patterns (e.g., developer coding conventions or the use of error-prone APIs) \cite{bosu2014identifying}. Existing passive frameworks (e.g., IRIS) terminate their process after verifying SAST alerts, thereby overlooking the rich pattern fingerprints \cite{kharkar2022learning} embedded within confirmed alerts (be they TPs or FPs). A truly intelligent system must possess the capacity to learn and generalize. It must be able to distill insights from verification results, generate new audit hypotheses, and start from known points to proactively excavate unknown, analogous vulnerabilities in the codebase caused by similar patterns.

Based on these insights, the overall design of vEcho aims to simulate the complete, end-to-end workflow of an experienced security expert auditing code. This workflow encompasses three primary phases.
\begin{itemize}
    \item \textbf{Initial Alerts.} Receiving and processing initial leads (e.g., SAST alerts).
    \item \textbf{Deep Verification.} Conducting verification that is deeply integrated with project-level context.
    \item \textbf{Proactive Discovery.} Learning from verified experiences and proactively generalizing these patterns to discover unknown threats.
\end{itemize}

The vEcho framework is methodologically language-agnostic. In this paper, we have instantiated it for Java Web vulnerability auditing. vEcho's workflow is divided into four interconnected logical stages, as illustrated in Figure \ref{fig:framework}, which we will elaborate on in Sections 3.2 through 3.5.

\subsection{Stage 1: Candidate Generation \& Filtering}
The objective of this stage is to provide the most comprehensive and relevant initial leads possible for the subsequent deep analysis. To achieve this, we employ a multi-source scanning strategy \cite{li2018vuldeepecker}\cite{li2021sysevr}. We run traditional, rule-based SAST tools (e.g., CodeQL and Coverity) in parallel to obtain a broad spectrum of alerts. Concurrently, we introduce a self-developed LLM-based static scanning agent, which we term ScanAgent.

ScanAgent complements traditional SAST tools. It is not constrained by manually crafted taint specifications; instead, it leverages the LLM's semantic understanding of code, aiming to capture more subtle candidate vulnerabilities that traditional rules might miss.

After scanning is complete, we perform two layers of automated filtering: first, deduplication and merging to consolidate highly similar alerts from different tools (CodeQL, Coverity, and ScanAgent); second, preliminary filtering to remove categories clearly unrelated to security risks (e.g., test code, dead code issues). This allows the candidates to converge on high-value security issues, providing a refined input set for the next stage (Section 3.3) of deep verification.

\subsection{Stage 2: Context-Aware Deep Verification}
This is vEcho's vulnerability alert verification phase, designed to address the lack of global project context limitation mentioned in Insight 1. For each candidate alert, vEcho does not immediately render a Yes/No judgment. Instead, it activates a deep verification agent we call VerifyAgent, which acts as the commander driving a structured, multi-turn analysis process.

During this process, vEcho actively and strategically dispatches its Tool Suite to gather evidence to answer a series of audit questions critical for vulnerability judgment. For example:
\begin{itemize}
    \item Invoking the Code Navigation Tool. Does the tainted data pass through effective sanitation or validation before reaching the sink (dangerous operation)?
    \item Invoking the Project Comprehension Tool. Is this API endpoint protected by project-level global filters or authentication (e.g., Spring Security)?
    \item Invoking the Web Search Tool. Are there known security advisories for the version of the third-party library being used?
\end{itemize}

Simultaneously, vEcho queries the Cognitive Memory Module (detailed in Section 3.4) to retrieve general security patterns and the current project's specific logic (e.g., a known internal sanitation function). Through this integration of tool invocation and information synthesis, vEcho constructs a complete, dynamic contextual view of the alert. This simulates the human expert's process of making a comprehensive judgment by combining code details, project architecture, and external knowledge, ultimately delivering a far more precise and interpretable judgment than simple classification. This judgment (whether TP or FP) is then passed as a fingerprint to the subsequent stages.

\subsection{Stage 3: Cognitive Learning \& Feedback}
To address the stateless limitation of existing work, we designed the Cognitive Memory Module, enabling vEcho to learn and grow from its audit experiences. This module is a key differentiator between vEcho and stateless analysis frameworks like IRIS. It achieves knowledge persistence and transfer, providing bidirectional support for both Deep Verification (Section 3.3) and Proactive Discovery (Section 3.5). The module consists of two main components.
\begin{itemize}
    \item \textbf{Temporary KB:} Equivalent to human working memory. It stores project-specific contextual information for the currently analyzed project, such as macro-architecture and core components.
    \item \textbf{Permanent KB:} Equivalent to human long-term memory. It stores cross-project, generalizable security knowledge and vulnerability patterns (e.g., the hazardous usage of a specific library, or a verified benign coding pattern).
\end{itemize}

Most critically, it incorporates a feedback learning loop. After vEcho confirms a vulnerability or a false positive, the system attempts to distill and generalize the universally applicable patterns or insights from that case. This new knowledge is then stored in the Permanent KB, allowing its analytical capabilities to continuously improve over time.

\subsection{Stage 4: Echoic Vulnerability Propagation}
This module is the key to vEcho's breakthrough from passive verification (1) to proactive discovery (N), and it embodies the core Echo metaphor. It is designed to address the SOTA limitation of capability boundary limited to verification \cite{cao2025recurring}. This module is driven by the output of the verification and includes two parallel trigger mechanisms.

\subsubsection*{Re-Scan Guidance Driven} Following every vEcho verification (whether True Positive or False Positive), vEcho leverages the cognitive memory capability from to reflect on the analysis and generate a set of executable Re-Scan Guidance. These guides are new auditing hypotheses that vEcho formulates based on its deep analysis. For example, after analyzing a false positive: This FP was due to the tool failing to recognize the actual functionality of FSTSerializer. Future scans should verify if serialized classes truly implement functionality and check control flow to confirm they are actually invoked.
    
\subsubsection*{Similarity Driven} This mechanism is activated after confirming a True Positive. It is based on the understanding that similar code stems from similar origins and hunts for accomplices throughout the codebase. This involves, on one hand, semantic-logic generalization, such as finding code that calls the same dangerous functions or exhibits similar data-flow patterns, and, on the other hand, developer-behavior generalization, such as finding code with similar functionalities.

\subsubsection*{EVP Execution and the Intelligent Loop} The outputs generated by these two mechanisms (i.e., new suspicious code and Re-Scan Guidance) are not sent directly to the verification module (Section 3.3). Instead, they are routed to our ScanAgent to initiate a secondary scan. 

We deliberately choose not to reuse CodeQL or Coverity at this stage. Executing exploratory tasks based on Re-Scan Guidance (e.g., scan all admin commands) or similarity requires a flexible scanner that does not rely on predefined rules \cite{bohme2017directed}. To this end, the EVP's secondary scan task is delegated to our self-developed ScanAgent (introduced in Section 3.2). As mentioned, ScanAgent's flexibility and independence from predefined rules make it the ideal choice for executing such exploratory tasks.

Furthermore, this design achieves Context Isolation architecturally: the VerifyAgent focuses on the deep, rigorous Verification context, while the ScanAgent focuses on the flexible, exploratory Scanning context.

Finally, the new suspicious code generated by this secondary scan is fed back into the Deduplication \& Merging queue of Stage 1 (Candidate Generation \& Filtering), forming a complete \textbf{Analyze} $\to$ \textbf{Learn} $\to$ \textbf{Rediscover} $\to$ \textbf{Re-analyze} intelligent closed loop.

\section{Evaluation}
This section aims to quantitatively validate the effectiveness of vEcho through a series of rigorous experiments. We elucidate our experimental setup in Section 4.1, including RQs, datasets, baselines, and evaluation metrics. Subsequently, we answer the research questions sequentially: We first analyze the impact of base models in Section 4.2 (RQ1); next, in Section 4.3, we conduct an end-to-end effectiveness comparison of vEcho against SOTA baselines (RQ2); then, in Section 4.4, we quantify the contributions of vEcho's core components through ablation studies (RQ3). Finally, to intuitively demonstrate the proactive discovery capability of the EVP mechanism (proven critical in RQ3), we provide a case study of a real-world 0-day vulnerability discovery in Section 4.5.

\subsection{Experimental Setup}
\subsubsection{Research Questions (RQs)}
Our evaluation aims to answer the following three core research questions:
\begin{itemize}
    \item \textbf{RQ1 (Impact of Base Model):} To what extent is the effectiveness of vEcho influenced by its underlying base LLM?
    \item \textbf{RQ2 (Effectiveness Comparison):} How does vEcho perform in terms of detection rate, Avg FDR, and Avg F1 score on vulnerability verification tasks compared to baseline methods like CodeQL, Coverity, IRIS, and Vanilla LLM?
    \item \textbf{RQ3 (Component Ablation Study):} What are the respective contributions of vEcho's core innovative components—the Cognitive Memory system and the EVP mechanism—to the overall performance (both in accuracy improvement and new vulnerability discovery)?
\end{itemize}

\subsubsection{Benchmark Dataset and Evaluation Validity}
To conduct the most direct and fair comparison against our core SOTA baseline (IRIS), we adopted the CWE-Bench-Java dataset \cite{iris_sast_cwe_bench_java}, which was constructed and publicly released in the IRIS study. This dataset comprises 120 manually verified Java projects containing real-world vulnerabilities, covering four common CWE types (CWE-22, CWE-78, CWE-79, CWE-94).

\subsubsection*{Anonymization for Evaluation Validity} The deep verification stage of vEcho (Section 3.3) is equipped with a web search tool. To prevent vEcho from gaining \textit{prior knowledge} through information leakage (e.g., by directly searching project names or CVE numbers), we performed a strict anonymization of the CWE-Bench-Java dataset. This process involved removing all original project names, Git history (\texttt{.git} information), dependency names in \texttt{pom.xml}, and any metadata within code comments that could reveal the vulnerability's identity (such as CVE numbers).

\subsubsection*{Handling and Extension of Ground Truth} We strictly use the 120 known vulnerabilities originally defined by CWE-Bench-Java as the uniform baseline for calculating detection rates. However, during our analysis, vEcho discovered 37 additional vulnerabilities not recorded in the original dataset but manually verified by us as true positives. To ensure fairness and accuracy in our evaluation, we adopted the following approach:
\begin{itemize}
    \item These newly discovered true vulnerabilities are \textit{not} counted towards the detection rate, ensuring a like-for-like comparison with the IRIS baseline on the 120-vulnerability benchmark.
    \item To maintain the strictest, most direct like-for-like comparison with IRIS, these 37 newly discovered true vulnerabilities were intentionally treated as False Positives when calculating the Avg FDR.
\end{itemize}

We must emphasize that the Avg FDR metric, consequently, significantly and artificially penalizes vEcho's proactive discovery capabilities. Therefore, these 37 vulnerabilities will be separately accounted for and discussed in our main analysis (Section 4.3) as direct proof of vEcho's Beyond-the-Benchmark discovery capabilities.

\subsubsection{Baseline Methods}
To comprehensively evaluate vEcho's performance, we compare it against the following three baselines, representing different methodologies:
\begin{itemize}
    \item \textbf{SAST Tools Only:} The raw scan results from CodeQL and Coverity. This baseline represents current standard industry practice, where both input and output originate from traditional SAST.
    \item \textbf{IRIS:} We strictly follow the methodology and configuration described in their paper \cite{li2024iris} and directly cite their paper's performance metrics as the SOTA benchmark to ensure the fairest comparison.
    \item \textbf{Vanilla LLM:} Alerts that passed the initial filtering (from Section 3.2) are fed directly to an LLM not equipped with any of vEcho's toolsets for judgment. This baseline aims to isolate and measure the benchmark verification performance of a pure LLM without assistance.
\end{itemize}

\subsubsection{Implementation Details and Metrics}

To answer RQ1 (Impact of Base Model), we evaluated the vEcho framework's performance when equipped with five prominent LLMs: GPT-4.1\cite{openai_gpt4_1_blog2025}, DeepSeek-R1-0528\cite{deepseek_r1_0528_2025}, GLM-4.5\cite{z_ai_glm4.5_blog}, Qwen3 8B\cite{qwen_ai_qwen3_blog}, and Qwen3 30B\cite{qwen_ai_qwen3_blog}. We note that the SOTA baseline IRIS utilized the GPT-4-0125-Preview model in their work \cite{li2024iris}. Given that this specific API version is now deprecated, we selected GPT-4.1 to ensure our comparison is based on a comparable state-of-the-art LLM. This model is the closest subsequent version available in terms of performance and architecture to the original IRIS baseline. This decision aims to ensure that vEcho and the SOTA baseline are evaluated on a fair and reproducible (SOTA-level) LLM foundation \cite{10646663}. For all other experiments (RQ2, RQ3), we uniformly use the LLM proven to perform best in RQ1 as vEcho's default reasoning engine (i.e., the driving LLM for both VerifyAgent and ScanAgent).

\subsubsection*{Evaluation Inputs} To clearly evaluate the contributions of different components, our experiments utilize two distinct sets of inputs:
\begin{itemize}
    \item \textbf{Baseline SAST Input:} Contains only the raw alerts from CodeQL and Coverity.
    \item \textbf{vEcho Extended Input:} Contains alerts from CodeQL, Coverity, and additional alerts discovered by our self-developed ScanAgent.
\end{itemize}

Our baseline methods and vEcho utilize these inputs as follows:
\begin{itemize}
    \item \textbf{SAST Tools Only:} Directly evaluates their raw output on the Baseline SAST Input.
    \item \textbf{IRIS:} We adopt the results reported in their paper.
    \item \textbf{Vanilla LLM:} To fairly disentangle the contributions of vEcho's Context-Aware Deep Verification (Section 3.3) and EVP mechanisms(Section 3.5), the Vanilla LLM uses the exact same vEcho Extended Input as vEcho.
    \item \textbf{vEcho:} Uses the vEcho Extended Input.
\end{itemize}

\subsubsection*{Evaluation Metrics} To ensure our evaluation results have the highest comparability with the SOTA baseline, we adopted the evaluation metrics defined in the IRIS\cite{li2024iris}.

We assume a dataset $\mathcal{D} = \{P_1, \dots, P_n\}$, where each $P_i$ is a Java project known to contain at least one vulnerability.
The label (Ground Truth) for a project $P$ is provided as a set of crucial program points $\mathbf{V}_{\text{vul}}^P = \{V_1, \dots, V_m\}$,
which the vulnerable paths must pass through.
Let $\textit{Paths}^P$ be the set of detected paths for each project $P$.
We first define $\textit{VulPath}(P)$ as the number of correctly identified vulnerability paths in $P$:

\begin{align}
\textit{VulPath}(P) &= |\{\textit{Path} \in \textit{Paths}^P ~|~ \textit{Path} \cap \mathbf{V}_{\text{vul}}^P \neq \emptyset\}|
\end{align}

If $\textit{VulPath}(P) > 0$, we consider the vulnerability detected. We define the project-level recall $\textit{Rec}(P)$ as a binary metric indicating this detection:

\begin{align}
\textit{Rec}(P) &= \mathbbm{1}_{\textit{VulPath}(P) > 0}
\end{align}

The total number of detected vulnerabilities, $\textit{Detected}(\mathcal{D})$, is the sum of this metric across all projects:

\begin{align}
\textit{Detected}(\mathcal{D}) &= \textstyle\sum_{P \in \mathcal{D}} \textit{Rec}(P)
\end{align}

Finally, we define the project-level precision $\textit{Prec}(P)$. To robustly handle cases where a tool retrieves no paths ($|\textit{Paths}^P| = 0$) and avoid division-by-zero, we explicitly define $\textit{Prec}(P)$ as 0 in such instances:

\begin{align}
\textit{Prec}(P) &= \begin{cases} \frac{\textit{VulPath}(P)}{|\textit{Paths}^P|} & \text{if } |\textit{Paths}^P| > 0 \\ 0 & \text{if } |\textit{Paths}^P| = 0 \end{cases}
\label{eq:4}
\end{align}

Based on \Cref{eq:4}, we calculate the final metrics. For $\textit{AvgFDR}$, a lower value is preferable. As noted in \cite{li2024iris}, to ensure $\textit{AvgFDR}$ is meaningful, we only average over projects where the tool produced at least one finding ($|\textit{Paths}^P| > 0$):

\begin{align}
\textit{AvgFDR}(\mathcal{D}) &= \operatorname*{avg}_{P \in \mathcal{D}, |\textit{Paths}^P|>0} (1 - \textit{Prec}(P))
\end{align}

For $\textit{AvgF1}$, we first define the project-level F1-score, $\textit{F1}(P)$. We again explicitly define $\textit{F1}(P)$ as 0 if both precision and recall are 0, preventing division-by-zero in the denominator:

\begin{align}
\textit{F1}(P) &= 
  \begin{cases} 
    \frac{2 \cdot \textit{Prec}(P) \cdot \textit{Rec}(P)}{\textit{Prec}(P) + \textit{Rec}(P)} & \substack{\text{if } (\textit{Prec}(P) \\ \qquad + \textit{Rec}(P)) > 0} \\[2ex]
    0 & \substack{\text{if } (\textit{Prec}(P) \\ \qquad + \textit{Rec}(P)) = 0}
  \end{cases} 
\\
\textit{AvgF1}(\mathcal{D}) &= \textstyle\frac{1}{|\mathcal{D}|} \textstyle\sum_{P \in \mathcal{D}} \textit{F1}(P)
\end{align}

\subsection{RQ1: Base Model Selection Analysis}

\begin{table*}
    \footnotesize
    \caption{
        Base LLM Selection (RQ1).
        This table shows the comparative end-to-end performance of vEcho using different LLMs on a 10-project subset. The results demonstrate the impact of LLM reasoning capacity and are used to select the optimal model for subsequent evaluations (RQ2 and RQ3).
        Performance is evaluated using Detection Rate ($\uparrow$ ), Avg FDR ($\downarrow$ ), and Avg F1 Score ($\uparrow$).
        \label{tab:baseModel}
        }
    \centering

    \setlength{\tabcolsep}{6pt}
    \begin{tabular}{rl|cccc}
        \toprule
        & \textbf{Method} & 
        \textbf{\#Detected} (/10) & \textbf{Detection Rate} ($\%$) &
        \textbf{Avg FDR} ($\%$) &
        \textbf{Avg F1 Score}
        \\
        \midrule
            \multirow{4}{*}{\textbf{vEcho} +} & 
            GPT-4.1 & 
            $\mathbf{6}$&
            $\mathbf{60}$& 
            $\mathbf{52.94}$& 
            $\mathbf{0.420}$
        \\
         & 
            DeepSeek-R1-0528 & 
            $5$ \red{($\downarrow 1$)} &
            $50$ \red{($\downarrow 10$)} & 
            $62.53$ \red{($\uparrow 9.59$)} & 
            $0.417$ \red{($\downarrow 0.003$)}
        \\
         &
            GLM-4.5 & 
            $6$ ($= 0$) &
            $60$ ($= 0$) &
            $70.97$ \red{($\uparrow 18.03$)} &
            $0.325$ \red{($\downarrow 0.095$)}
        \\
         & 
            Qwen3 30B & 
            $4$ \red{($\downarrow 2$)} &
            $40$ \red{($\downarrow 20$)} &
            $89.47$ \red{($\uparrow 36.53$)} &
            $0.191$ \red{($\downarrow 0.229$)}
        \\
        &
            Qwen3 8B & 
            $6$ ($= 0$)& 
            $60$ ($= 0$) &
            $86.96$ \red{($\uparrow 34.02$)} & 
            $0.222$ \red{($\downarrow 0.198$)}
        \\
        \bottomrule
    \end{tabular}
    \vspace{-8px}

\end{table*}

This research question (RQ1) aims to evaluate the extent to which the vEcho framework's performance is affected by the reasoning capabilities of its underlying base LLM. Since vEcho's Deep Verification (Section 3.3) and EVP mechanism (Section 3.5) both rely heavily on the LLM's contextual understanding and logical reasoning, selecting a base model with a balance of performance and cost is crucial.

Due to the high cost of LLM API calls, we used a randomly sampled subset of 10 projects from the CWE-Bench-Java dataset for this experiment. We evaluated the vEcho framework equipped with four different LLMs on this subset. We assessed the core metrics identical to those in Section 4.1.4: Detected, Detection Rate, Avg FDR, and Avg F1 Score.

\subsubsection*{Analysis and Selection} The results, presented in Table 1, reveal a clear performance stratification among the models. As our evaluation prioritizes the reduction of Alert Fatigue, the Avg FDR is our primary metric of concern.

GPT-4.1 clearly achieved the best overall performance. It secured the lowest Avg FDR ($52.94\%$) while also attaining the joint-highest detection rate ($60\%$), resulting in the top Avg F1 Score ($0.420$). Notably, DeepSeek-R1-0528 demonstrated highly competitive performance; despite a slightly lower detection count, its Avg F1 Score of $0.417$ is nearly identical to that of GPT-4.1.

Conversely, while GLM 4.5 and Qwen3 8B also achieved a $60\%$ detection rate, their utility is severely undermined by substantially higher false discovery rates ($70.97\%$ and $86.96\%$, respectively). This aligns with our qualitative observation that these models, likely due to smaller parameter counts, lack the sophisticated code understanding required for deep verification. They exhibit a tendency to classify alerts as true positives, inflating detection at the cost of precision. The Qwen3 30B model performed the poorest overall, with the lowest detection rate ($20\%$) and the highest Avg FDR ($89.47\%$), yielding the worst Avg F1 Score ($0.191$).To pose the strongest challenge to the SOTA (IRIS) and evaluate vEcho's peak performance in subsequent experiments (RQ2 and RQ3), we select GPT-4.1 as the default model for vEcho in all following experiments.

\begin{table*}
    \footnotesize
    \caption{
    End-to-End Performance Comparison Against SOTA Baselines (RQ2).
    Results are benchmarked on the CWE-Bench-Java dataset (120 vulnerabilities), comparing vEcho against three baselines: 
    IRIS (SOTA LLM-assisted), SATS Tools (industrial practice), and Vanilla LLM (unassisted).
    \label{tab:rq2-performance}
    \label{tab:performance}
        }
    \centering
    \setlength{\tabcolsep}{7pt}
    \begin{tabular}{rl|cccc}
        \toprule
        & \textbf{Method} & 
        \textbf{\#Detected} (/120) & \textbf{Detection Rate} ($\%$) &
        \textbf{Avg FDR} ($\%$) &
        \textbf{Avg F1 Score} 
        \\
        \midrule

        & 
            vEcho & 
            $\mathbf{78}$  &
            $\mathbf{65.00}$ & 
            $\mathbf{59.78}$ & 
            $\mathbf{0.422}$ 
        \\
         & 
            IRIS & 
            $55$ \red{($\downarrow 23$)} &
            $45.83$ \red{($\downarrow 19.17$)} & 
            $84.82$ \red{($\uparrow 25.04$)} & 
            $0.177$ \red{($\downarrow 0.245$)} 
        \\
         & 
            SATS Tools & 
            $29$ \red{($\downarrow 49$)} &
            $24.17$ \red{($\downarrow 40.83$)} & 
            $94.16$ \red{($\uparrow 34.38$)} & 
            $0.063$ \red{($\downarrow 0.359$)} 
        \\
         & 
            Vanilla LLM & 
            $34$ \red{($\downarrow 44$)} &
            $28.33$ \red{($\downarrow 36.67$)} & 
            $86.45$ \red{($\uparrow 26.67$)} & 
            $0.220$ \red{($\downarrow 0.222$)} 
        \\
        \bottomrule
    \end{tabular}
    \vspace{-3px}

\end{table*}

\subsection{RQ2: End-to-End Effectiveness Comparison}
This research question (RQ2) aims to directly compare vEcho as a complete, end-to-end audit system against the SOTA baseline (IRIS), pure LLM capabilities (Vanilla LLM), and traditional SAST tools. We ran vEcho and all baseline methods on the full CWE-Bench-Java dataset (120 projects). As established in Section 4.2, vEcho and Vanilla LLM both use GPT-4.1 as their reasoning engine.

vEcho demonstrates high precision in filtering false positives. As shown in \Cref{tab:performance}, vEcho achieves an Avg FDR ($59.78\%$). This result marks a substantial reduction in false positives compared to both the SOTA baseline IRIS (($84.82\%$)) and the Vanilla LLM ($86.45\%$).

It is noteworthy that the Vanilla LLM achieves a higher Avg F1 Score ($0.220$) than the SOTA baseline IRIS ($0.177$). We attribute this phenomenon primarily to vEcho's Candidate Generation And Filtering stage (see vEcho Extended Input in Section 4.1.4). This preprocessing stage effectively filters out a massive volume of low-quality alerts (such as code style or quality issues), thereby significantly improving the quality of the candidate set fed to the LLM.

In contrast, the IRIS baseline methodology tends to treat all alerts as candidates for verification without effective pre-filtering. This results in its candidate set being inundated with numerous entries unrelated to vulnerabilities, which severely impacts its Avg F1 score. This Avg F1 score, by definition, is significantly penalized when the same number of true positives must be identified within a massive pool of irrelevant alerts (e.g., 2 TPs in 100 alerts) compared to a high-quality, refined set (e.g., 2 TPs in 10 alerts). Therefore, this comparison indirectly but strongly corroborates the necessity and effectiveness of vEcho's Candidate Generation \& Filtering stage, which lays a solid foundation for subsequent vulnerability analysis by refining the candidate set.

\subsubsection*{Proactive discovery capabilities significantly boost recall} vEcho also substantially surpasses all baselines in Detection Rate, achieving a Detection Rate ($65.00\%$, or 78/120) over IRIS's ($45.83\%$, or 55/120). This improvement is by design and is directly attributable to two of vEcho's unique innovations, which are absent in the baselines: the Cognitive Memory Module, which provides project-level context, and the EVP mechanism, which dynamically discovers new candidates during the audit. 

\begin{figure}[t] %
    \centering
    \includegraphics[width=\columnwidth]{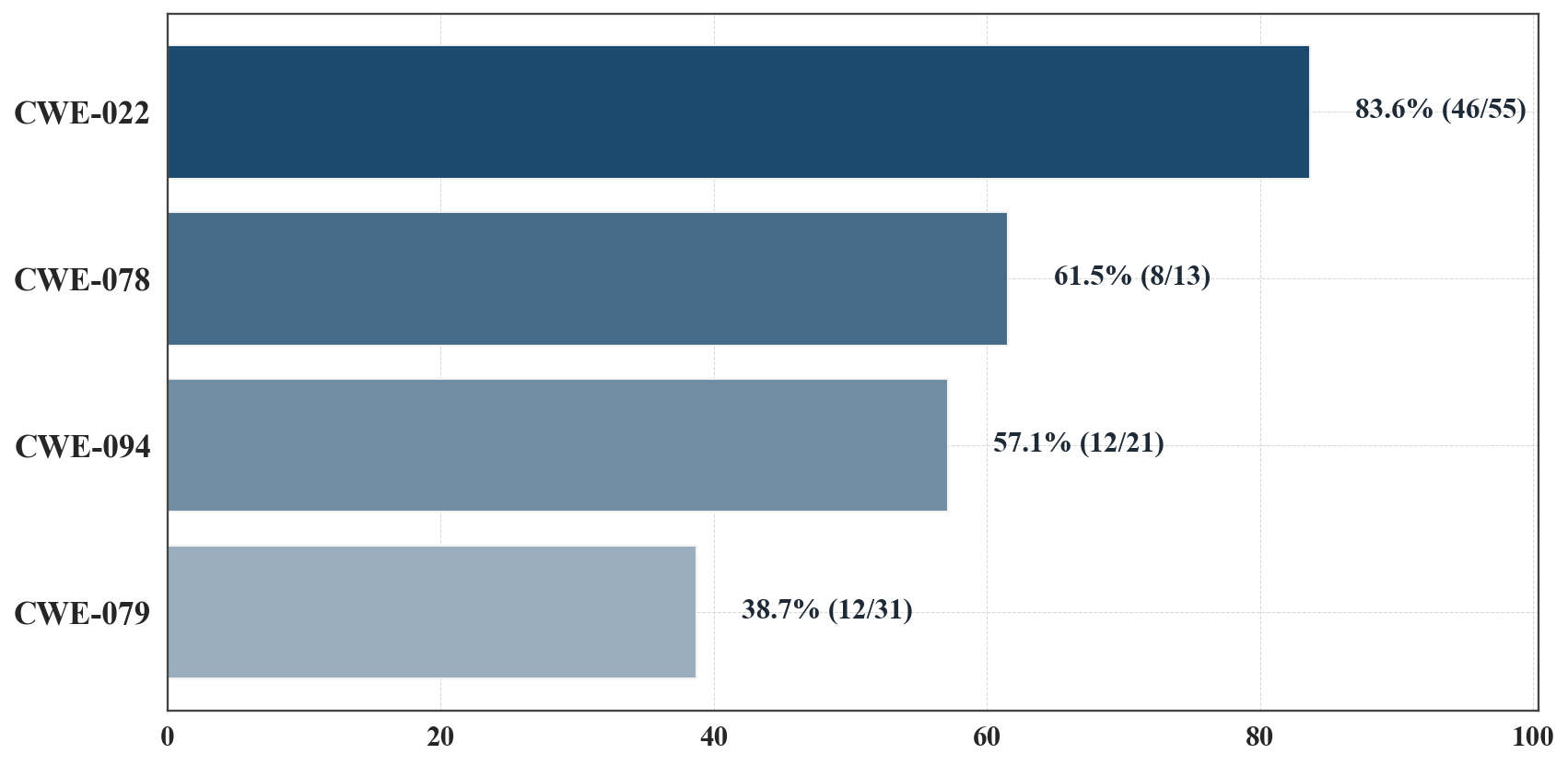} 
    \caption{Breakdown of vEcho's detection rate by CWE type on the CWE-Bench-Java dataset. The percentages and counts (e.g., 83.6\% (46/55)) match the data in Section 4.3.}
    \label{fig:verifyed_cve}
\end{figure}

As illustrated in Figure \ref{fig:verifyed_cve}, this detection is comprehensive, with high performance in critical categories such as CWE-022 (Path Traversal) at $83.6\%$ (46/55).

\subsubsection*{The most compelling metric is New Vulnerability Discoveries} vEcho's EVP mechanism proactively discovered 37 vulnerabilities \textit{not marked by the benchmark's ground truth}. To be clear, these 37 are true vulnerabilities found within the CWE-Bench-Java dataset but were not labeled by the original benchmark. 

\begin{figure}[t] 
    \centering
    \includegraphics[width=\columnwidth]{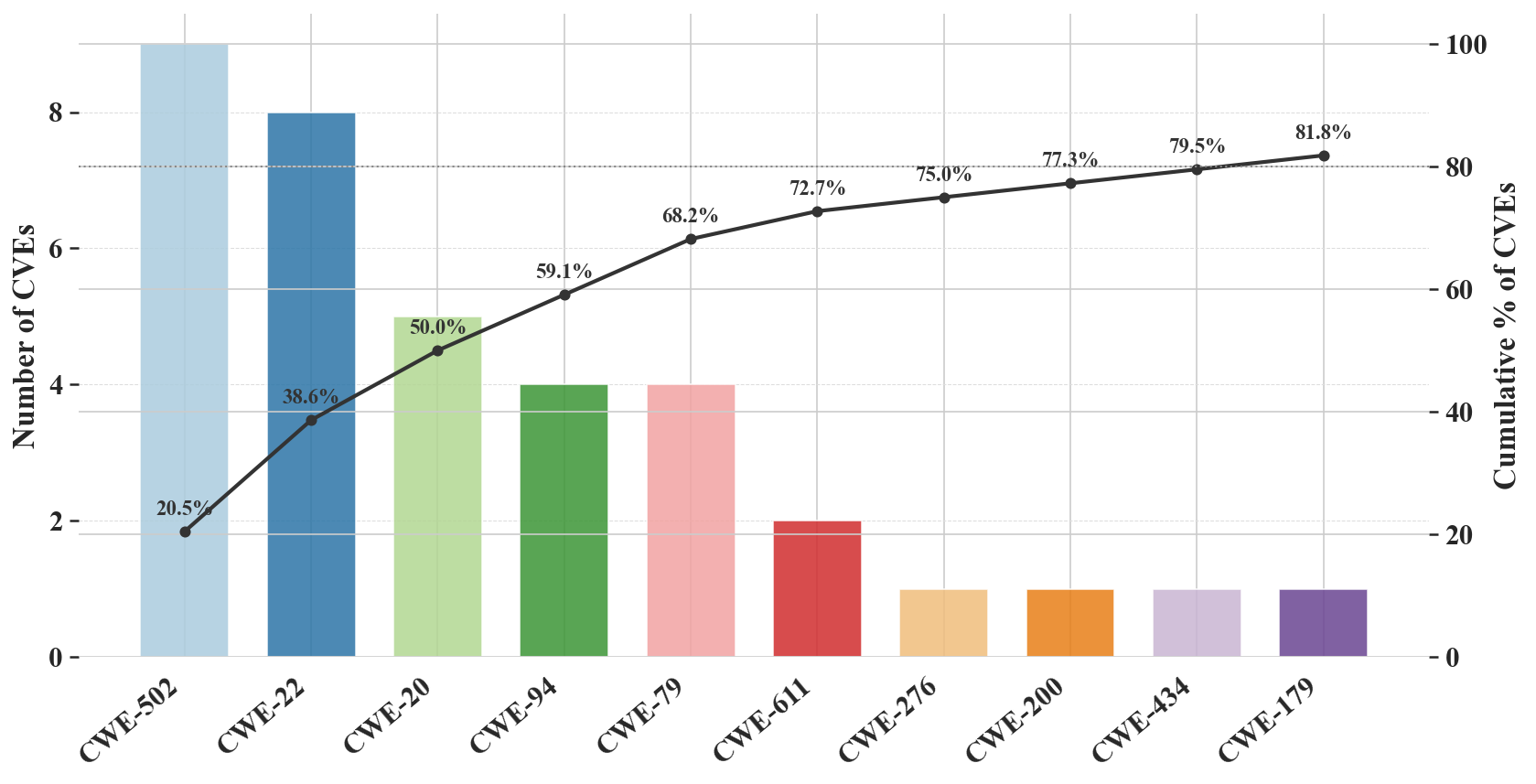} 
    \caption{Breakdown of the 37 Beyond-the-Benchmark vulnerabilities discovered by vEcho within the CWE-Bench-Java dataset. The chart shows the distribution by CWE type, highlighting a concentration in critical vulnerability classes.}
    \label{fig:knowncve2}
\end{figure}

Figure \ref{fig:knowncve2} provides a detailed breakdown of these discoveries, illustrating that they are not trivial, but rather concentrated in high-impact CWE categories such as Deserialization (CWE-502) and Path Traversal (CWE-22).

This clearly demonstrates that vEcho has achieved a qualitative leap from passive verification to proactive discovery. We will further demonstrate vEcho's ability to find novel 0-day vulnerabilities when auditing other real-world projects in the case study in Section 4.5.

\subsubsection*{Conclusion}vEcho comprehensively and significantly surpasses the existing SOTA baseline in Detection Rate, Avg FDR, and Avg F1 Score. The experimental data proves that this leap is thanks to vEcho's unique virtual security expert framework, which addresses Alert Fatigue through project-level context awareness and achieves true proactive discovery via the EVP mechanism.

\subsection{RQ3: Component Contribution (Ablation Study)}
In RQ2, we demonstrated that vEcho's full framework provides a massive performance leap over the Vanilla LLM, with the Avg F1 Score increasing from $0.220$ to $0.422$ (\Cref{tab:ablation}). This research question (RQ3) aims to use an Ablation Study to further quantify the respective contributions of vEcho's two core innovations—the Developer Tool Suite \& Cognitive Memory (Sections 3.3, 3.4) and the EVP Mechanism (Section 3.5)—to this performance improvement.

\begin{table*}
        \footnotesize
        \caption{
            Ablation Study of vEcho's Core Components. 
            This table isolates the performance contribution of the EVP mechanism (Section 3.5) and the Cognitive Memory / Developer Tools (Sections 3.3, 3.4). All metrics are evaluated on the CWE-Bench-Java dataset. 
            \label{tab:ablation} 
        }
        \centering
        \setlength{\tabcolsep}{6pt}
        \begin{tabular}{rl|cccc}
        \toprule
        & \textbf{Method} &
        \textbf{\#Detected} (/120) & \textbf{Detection Rate} ($\%$) &
        \textbf{Avg FDR} ($\%$) &
        \textbf{Avg F1 Score}
        \\
        \midrule
        &
            vEcho (Full System) &
            $\mathbf{78}$ &
            $\mathbf{65.00}$ &
            $\mathbf{59.78}$ &
            $\mathbf{0.422}$
        \\
        \midrule
        &
            vEcho w/o EVP &
            $62$ \red{($\downarrow 16$)} & %
            $51.67$ \red{($\downarrow 13.33$)} &
            $55.51$ \green{($\downarrow 4.72$)} & 
            $0.370$ \red{($\downarrow 0.052$)}
        \\
         &
            vEcho w/o Tools \& Memory &
            $49$ \red{($\downarrow 29$)} &
            $40.83$ \red{($\downarrow 24.17$)} &
            $84.97$ \red{($\uparrow 25.19$)} &
            $0.271$ \red{($\downarrow 0.151$)}
        \\
        &
            vEcho w/o both (Vanilla LLM) &
            $34$ \red{($\downarrow 44$)} &
            $28.33$ \red{($\downarrow 36.67$)} &
            $86.45$ \red{($\uparrow 26.67$)} &
            $0.220$ \red{($\downarrow 0.222$)}
        \\
        \bottomrule
    \end{tabular}
    \vspace{-8px}

\end{table*}

We start from the vEcho (Full System) configuration and progressively remove the key components, evaluating the end-to-end performance on the full CWE-Bench-Java dataset.
\begin{itemize}
    \item \textbf{vEcho w/o EVP:} We disabled all proactive discovery functions (EVP) from Section 3.5.
    \item \textbf{vEcho w/o Cognitive Memory:} We disabled the developer tool suite and cognitive memory described in Sections 3.3 and 3.4. In this configuration, the vEcho agent cannot actively investigate code context or learn from memory, forcing it to make zero-shot judgments on alerts.
    \item \textbf{vEcho w/o All (Vanilla LLM):} This baseline (from RQ2) serves as the full ablation baseline; it has no tool suite, cognitive memory, or EVP.
\end{itemize}

\subsubsection*{Analysis of Results} \Cref{tab:ablation} clearly illustrates that the contributions of vEcho's two core innovative components are orthogonal and complementary.

\subsubsection*{Developer Tool Suite \& Cognitive Memory} Removing this component causes the Avg F1 Score to drop sharply from $0.422$ to $0.271$. More critically, for our goal of mitigating alert fatigue, the Avg FDR \textbf{skyrockets from $59.78\%$ to $84.97\%$}, a 25.19 percentage point gap that leaves it almost as high as the Vanilla LLM ($86.45\%$). This 25.19 percentage point gap perfectly quantifies the superiority of active investigation over passive judgment. This result strongly proves that the reason vEcho's Deep Verification framework (Section 3.3) achieves such significant false positive filtering is its ability to actively invoke the tool suite and incorporate Cognitive Memory to acquire project-level context. Without this component, vEcho's verification capability degrades substantially.
\subsubsection*{EVP Mechanism} After removing the EVP mechanism, the \textbf{number of detected vulnerabilities ($\#Detected$) drops from 78 to 62}, a net loss of 16 true vulnerabilities. This directly causes the Avg F1 Score to fall from $0.422$ to $0.370$. This result perfectly isolates and proves that the EVP mechanism is the core engine for vEcho's paradigm shift from passive verification to proactive discovery, significantly boosting the system's Recall.

\subsubsection*{Conclusion} vEcho's Developer Tool Suite / Cognitive Memory and EVP Mechanism are both indispensable for achieving its dual advantages high precision and high recall. The vEcho full framework (Avg F1 Score $0.422$) delivers a massive 20.2 percentage point performance gain over the fully ablated Vanilla LLM baseline (Avg F1 Score $0.220$), a 91.8\% relative improvement in Avg F1 score—which methodologically proves the superiority of our virtual security expert paradigm.

\subsection{Case Study: Proactive Discovery of a 0-day Vulnerability}
\begin{figure*}
    \centering
    \includegraphics[width=1\textwidth]{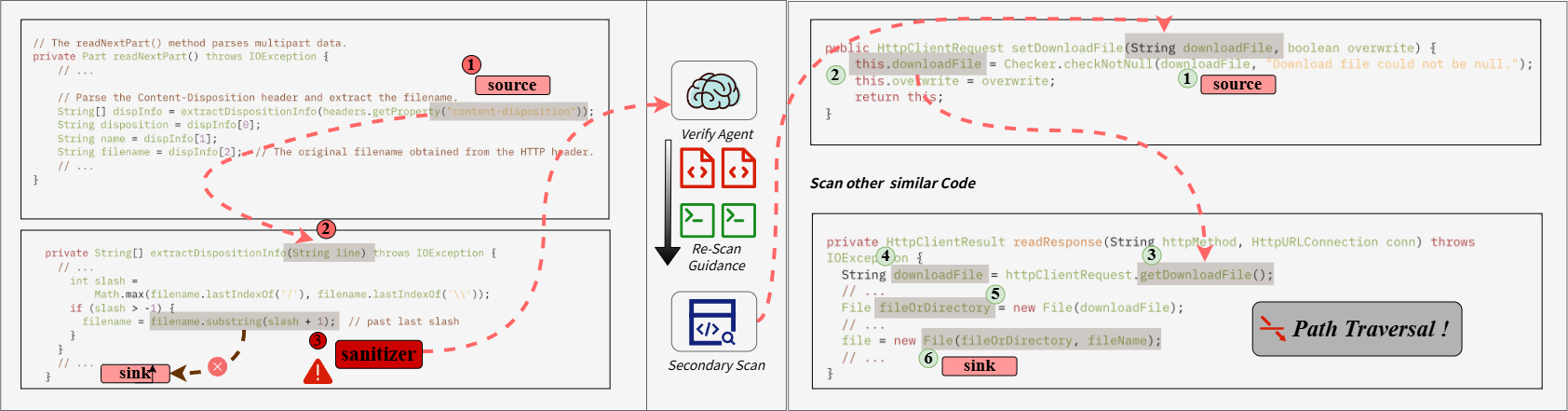}
    \caption{Case study of vEcho's 0-day discovery via Echoic Vulnerability Propagation (EVP). 
    (Left) vEcho analyzes a SAST alert in the Resty framework, where a \texttt{source} \textbf{(snippet 1)} is correctly neutralized by a project-specific \texttt{sanitizer} \textbf{(snippet 3)}, resulting in a False Positive (FP). (Middle) Instead of discarding the FP, the Verify Agent learns this sanitizer's pattern and generates 'Re-Scan Guidance'. (Right) The EVP mechanism uses this guidance to proactively discover a new, un-sanitized code path in the HttpClient framework, confirming a 0-day Path Traversal vulnerability at the \texttt{sink} \textbf{(snippet 6)}.}
    \label{fig:rescan}
\end{figure*}

To intuitively demonstrate the practical capabilities of the vEcho framework, especially its exceptional performance in discovering high-impact security vulnerabilities, we provide an in-depth analysis of a real-world 0-day vulnerability discovery process.

vEcho's effectiveness has been confirmed in audits of multiple open-source projects, totaling the discovery of 51 novel 0-day vulnerabilities. Critically, these discoveries include high-impact Critical Remote Code Execution (RCE) and Arbitrary Code Execution (ACE) vulnerabilities, which have been officially confirmed by the Apache project, fully demonstrating vEcho's real-world capabilities.

The case study in this section (\Cref{fig:rescan}) details the core mechanism by which vEcho achieves this: how its EVP mechanism (Section 3.5) learns from analysis experience (even from false positives) and ultimately realizes the Analyze $\to$ Learn $\to$ Rediscover intelligent closed loop.

\subsubsection*{Stage 1: False Positive Verification and Learning} vEcho's workflow begins with a file upload path traversal alert reported by a SAST tool (Section 3.2) in the Resty framework. During the context-aware deep verification stage (Section 3.3), the vEcho traces the taint \texttt{source} (\Cref{fig:rescan}, left, snippet 1) into the \texttt{extractDispositionInfo} method (snippet 2).

Code analysis confirms that this method's body contains a critical sanitization step (\Cref{fig:rescan}, left, snippet 3). It removes all potential path separators from the filename by truncating the string to the last occurring \texttt{/} or \texttt{$\setminus$}. Because this sanitizer prevents the taint from reaching the \texttt{sink}, vEcho intelligently determines the alert to be a False Positive (FP).

\subsubsection*{Stage 2: Generalization Trigger and Exploration}However, vEcho's workflow does not stop here. In a passive framework (like IRIS), this FP would simply be discarded. In vEcho, the cognitive memory module (Section 3.4) is activated.

vEcho reflects on this false positive analysis and distills a key insight: The SAST tool reported a false positive because it failed to recognize the logic in snippet 3 as a project-specific sanitizer. This insight points to a critical hypothesis:

\textit{“There may be other code paths in the codebase that handle \texttt{Content-Disposition} headers but are not protected by this sanitizer.”}

This insight is formalized into an executable Re-Scan Guidance (Section 3.5), with the core objective:

\textit{“Due to this false positive, initiate a search for other potential directory traversal points that are not filtered by this path processing function.”}

\subsubsection*{Stage 3: Proactive Discovery of 0-day}
This Re-Scan Guidance activates vEcho's EVP mechanism (Section 3.5). This Re-Scan Guidance activates vEcho's EVP mechanism (Section 3.5). The EVP mechanism follows this new directive, conducting a targeted search of the codebase for other code paths that process \texttt{Content-Disposition} headers.

The ScanAgent subsequently targets the HttpClient file download functionality within the project, a completely new code path not flagged by any SAST tool. This new candidate is sent back to the deep verification queue. Upon verification, vEcho discovers:

\subsubsection*{Identical Taint Source} Also retrieves \texttt{fileName} from \texttt{conn.getHeaderField("Content-Disposition")}.
    
\subsubsection*{Missing Sanitizer} Crucially, this code path does not call the \texttt{extractDispositionInfo} sanitizer. Instead, it uses a simple \texttt{extractDispositionInfo} call that fails to validate the path (\Cref{fig:rescan}, right, snippet 5).
    
\subsubsection*{Vulnerable Sink} This un-validated \texttt{fileName} is then passed directly to \texttt{new File(fileOrDirectory, fileName)} (\Cref{fig:rescan}, right, snippet 6), leading to a classic Path Traversal vulnerability\cite{jovanovic2006pixy}.

\subsubsection*{Case Conclusion} This case study clearly demonstrates how vEcho learns from the analysis of a false positive, generates intelligent Re-Scan Guidance, and proactively generalizes that pattern to discover a hidden, real-world 0-day vulnerability. This is a core capability that passive frameworks like IRIS are architecturally incapable of performing.

\section{Discussion}
This section first discusses the Limitations of the vEcho methodology and the corresponding Future Work. Subsequently, we will elaborate on the Ethics Considerations and the responsible disclosure process undertaken to handle the 0-day vulnerabilities discovered in this research.

\subsection{Limitations and Future Work}

Despite vEcho demonstrating significant performance advantages in the experimental evaluation, it has three primary limitations.

\subsubsection*{Reliance on the Base LLM's Reasoning Capability} vEcho's performance is deeply coupled with the reasoning capability of its virtual expert (i.e., the base LLM). Although our Deep Verification framework (Section 3.3) anchors the LLM's analysis with external evidence provided by the tool suite, inherent hallucinations or erroneous understanding of complex logic by the LLM can still lead to misjudgements \cite{liu2023trustworthy}\cite{tonmoy2024comprehensive}.
    
\subsubsection*{Implementation and Language Generality}vEcho's core framework (i.e., the Verification-Learning-EVP intelligent closed loop) is language-agnostic by design. However, our current instantiation is focused on the Java Web ecosystem. This limits vEcho's current direct applicability to other languages (such as Python or C/C++) or non-standard project structures.
    
\subsubsection*{Overhead and Scalability of Deep Analysis}vEcho's Deep Verification (Section 3.3) involves multi-turn LLM calls and tool executions. Although its significant precision improvement (as shown in RQ2 and RQ3, its Avg FDR is far lower than the baselines) justifies this overhead, its analysis overhead is markedly higher compared to traditional SAST. This may pose challenges when conducting full-scale audits of ultra-large-scale projects (e.g., the entire Linux kernel).

Corresponding to the limitations above, future research directions include:

\subsubsection*{Improving Reliability} Exploring the introduction of stronger formal logic constraints is a promising research direction to further reduce hallucinations and improve verification reliability \cite{pan-etal-2023-logic}\cite{morishita2024enhancing}\cite{lin2025zebralogic}.
    
\subsubsection*{Improving Scalability} Investigating more intelligent alert prioritization and incremental analysis strategies (e.g., only analyzing code that has changed since the last commit and its dependencies) is key to improving vEcho's analysis efficiency on large-scale projects \cite{9700318}\cite{gelman2023escalatedquicklymlframework}.
    
\subsubsection*{Extending Language Generality} A critical future direction is to extend vEcho's tool suite interface to support static analysis libraries for other languages (e.g., the \texttt{ast} library for Python or \texttt{Clang} for C/C++). This would validate the generality of vEcho's paradigm shift in broader ecosystems \cite{279967}\cite{10.1145/3468264.3468538}.

\subsection{Ethics Considerations}
The core objective of this research is to advance automated vulnerability discovery techniques to empower defenders and enhance the security of the software ecosystem. We are keenly aware of the dual-use risk \cite{zhang2025abusability}\cite{hasegawa2024weird} that such automated offensive research could be misused.

Therefore, we are committed to handling our findings responsibly. For all previously unknown 0-day vulnerabilities discovered during our experiments (including the critical RCE and ACE vulnerabilities found in Apache projects), we have strictly adhered to the Coordinated Vulnerability Disclosure (CVD) principles \cite{first2020cvd}. We have privately reported them to the corresponding project maintainers or organizations through secure, official channels at the earliest opportunity.

In this paper, discussions of these vulnerabilities have been desensitized or obscured, and are only included after receiving permission for remediation or disclosure. We will not publicly release any unrestricted tools or data that could be directly used for large-scale, automated attacks.

\section{Conclusion}
Addressing the fundamental deficiencies of existing LLM-assisted vulnerability analysis methods—which are passive, stateless, and confined to isolated contexts—this paper introduces vEcho, a novel agent framework.

Unlike SOTA methods such as IRIS, which aim to optimize the SAST process, vEcho achieves a paradigm shift. By equipping its core LLM with a comprehensive developer tool suite and an innovative Cognitive Memory Module, vEcho successfully transforms the LLM from a passive classifier into a virtual security expert. This expert is capable of simulating human specialists and performing project-level context-aware deep audits.

More importantly, our novel Echoic Vulnerability Propagation (EVP) mechanism enables vEcho to learn from both successful (TPs) and failed (FPs) verifications, generate intelligent Re-Scan Guidance, and proactively discover unknown, an alogous vulnerabilities. This achieves a significant capability breakthrough from passive verification to proactive discovery.

Extensive experiments on the CWE-Bench-Java benchmark, coupled with the practical discovery of 51 novel 0-day vulnerabilities in real-world projects (including high-impact Apache RCE/ACE flaws), demonstrate vEcho's massive leap over the state-of-the-art. This is evidenced by its significantly superior Avg FDR compared to the SOTA baseline and its unparalleled proactive discovery capabilities.

Our work not only demonstrates a viable path to forging LLMs into virtual security experts but also offers a blueprint for constructing the next generation of AI-driven, automated security auditing systems truly capable of learning, memory, and reasoning.

\section*{LLM usage considerations}
In adherence to the IEEE S\&P policy on Large Language Models (LLMs), we disclose our use of LLMs in this research.

\subsection*{Originality} This manuscript, including all scientific claims, methodological designs, and experimental analyses, represents our original intellectual work. We are responsible for the entire content of this paper. We are also responsible for the thoroughness of our literature review, ensuring relevant prior work is properly credited.

Our core methodology, the vEcho framework, utilizes LLM as a central component. The framework's design, including the developer tool suite, the Cognitive Memory Module, and the Echoic Vulnerability Propagation (EVP) mechanism, is a novel contribution developed entirely by us. The LLM is employed as a reasoning engine within this framework, which we designed. All conclusions drawn from experimental results (Sections 4.2-4.5) were analyzed and validated by us. Additionally, LLMs were used for editorial purposes (e.g., grammar correction and style refinement) in this manuscript, and we manually inspected and revised all outputs to ensure accuracy and originality.

\subsection*{Transparency} 
\subsubsection*{LLM in Methodology} Our framework (vEcho) is explicitly designed to leverage the reasoning capabilities of LLMs, as detailed in Section 3. vEcho uses LLMs via proprietary APIs as its core inference engine.

\subsubsection*{Model Selection} We conducted experiments using several foundational models, as detailed in our RQ1 analysis (Section 4.2). We selected a GPT-4.1 (as the gpt-4-0125-preview used by IRIS is the history API) as the primary model for vEcho in our main evaluation. This decision was based on two factors: 
\begin{enumerate} 
    \item \textbf{State-of-the-Art Comparison.} Our primary baseline, IRIS, utilized a GPT-4 model (gpt-4-0125-preview) for its best results. The selected GPT-4.1 provides performance closest to the gpt-4-0125-preview among available models, ensuring a fair and robust comparison against the state-of-the-art baseline. 
    \item \textbf{Experimental Consistency and Cost.} Our extensive experimental campaign, covering 120 projects in CWE-Bench-Java and additional real-world projects, was a large-scale, longitudinal effort. Switching to a different model family mid-evaluation would have been cost-prohibitive and, more importantly, would have compromised the consistency of our analysis. We proceeded with the established, high-performance GPT-4.1model to ensure valid results. 
\end{enumerate}

\subsubsection*{Reproducibility} A key limitation introduced by using closed-source, non-deterministic API-based LLMs (like GPT-4.1) is the challenge of exact reproducibility. While our framework and its components will be open-sourced, results from the LLM component may vary slightly across API versions or over time. We mitigate this by using a fixed model version (GPT-4.1) during our evaluation. Furthermore, our Cognitive Memory System (Section 3.4), which will also be open-sourced, is explicitly designed to constrain the LLM's reasoning process. This promotes more convergent and consistent decisions, further mitigating the non-determinism of the API.

\subsection*{Responsibility} 
\subsubsection*{Data Ethics} This work did not involve training new LLMs. Our data collection was limited to publicly available, open-source code repositories (namely, the CWE-Bench-Java dataset and other public GitHub projects). All data analyzed was open-source, mitigating ethical concerns regarding data consent or intellectual property. All 0-day vulnerabilities discovered during this research were responsibly disclosed to the respective project maintainers following standard Coordinated Vulnerability Disclosure (CVD) protocols (Section 5.2).

\subsubsection*{Necessity of LLM} An LLM is fundamentally necessary for this research. Our core thesis is to shift the paradigm from LLM-assisted SAST to an LLM-driven virtual security expert. This requires capabilities for deep semantic understanding, learning, and human-like reasoning that traditional deep learning models cannot provide.

\subsubsection*{Model Size Selection} Our RQ1 analysis (Section 4.2) explicitly justifies the need for a high-parameter model. The results demonstrate that the complex, multi-step reasoning required for deep vulnerability analysis (Section 3.3) and Echoic Vulnerability Propagation (Section 3.5) is only achievable with large-scale models like GPT-4. Smaller models, while tested, did not provide the necessary reasoning quality to validate our framework's contributions.

\subsubsection*{Minimizing Queries} The vEcho framework is explicitly designed to minimize the environmental and economic footprint of LLM queries. Unlike naive approaches, vEcho uses: (1) Candidate Generation \& Filtering (Section 3.2) to heavily prune low-quality alerts before engaging the expensive LLM, and (2) the Cognitive Memory System (Section 3.4) to store and reuse knowledge, avoiding redundant analysis of similar code patterns. This ensures that high-cost LLM reasoning is reserved only for high-priority, novel, and context-rich candidates.

\subsubsection*{Hardware} The vEcho framework and all associated experiments were orchestrated on local  servers. This hardware (utilizing standard CPUs, RAM, and NVIDIA RTX 4090 GPUs) was used for all local tasks, including running smaller foundational models (e.g., Qwen3 8B) for our RQ1 analysis. The hardware specific to running the large-scale LLM (GPT-4.1) inference itself is managed by the proprietary API provider (e.g., OpenAI).

\bibliographystyle{IEEEtran}
\bibliography{references}

\begin{thebibliography}{10}
\providecommand{\url}[1]{#1}
\csname url@samestyle\endcsname
\providecommand{\newblock}{\relax}
\providecommand{\bibinfo}[2]{#2}
\providecommand{\BIBentrySTDinterwordspacing}{\spaceskip=0pt\relax}
\providecommand{\BIBentryALTinterwordstretchfactor}{4}
\providecommand{\BIBentryALTinterwordspacing}{\spaceskip=\fontdimen2\font plus
\BIBentryALTinterwordstretchfactor\fontdimen3\font minus \fontdimen4\font\relax}
\providecommand{\BIBforeignlanguage}[2]{{%
\expandafter\ifx\csname l@#1\endcsname\relax
\typeout{** WARNING: IEEEtran.bst: No hyphenation pattern has been}%
\typeout{** loaded for the language `#1'. Using the pattern for}%
\typeout{** the default language instead.}%
\else
\language=\csname l@#1\endcsname
\fi
#2}}
\providecommand{\BIBdecl}{\relax}
\BIBdecl

\bibitem{johnson2013don}
B.~Johnson, Y.~Song, E.~Murphy-Hill, and R.~Bowdidge, ``Why don't software developers use static analysis tools to find bugs?'' in \emph{2013 35th International Conference on Software Engineering (ICSE)}.\hskip 1em plus 0.5em minus 0.4em\relax IEEE, 2013, pp. 672--681.

\bibitem{pearce2023examining}
H.~Pearce, B.~Tan, B.~Ahmad, R.~Karri, and B.~Dolan-Gavitt, ``Examining zero-shot vulnerability repair with large language models,'' in \emph{2023 IEEE Symposium on Security and Privacy (SP)}.\hskip 1em plus 0.5em minus 0.4em\relax IEEE, 2023, pp. 2339--2356.

\bibitem{ullah2024llms}
S.~Ullah, M.~Han, S.~Pujar, H.~Pearce, A.~Coskun, and G.~Stringhini, ``Llms cannot reliably identify and reason about security vulnerabilities (yet?): A comprehensive evaluation, framework, and benchmarks,'' in \emph{2024 IEEE Symposium on Security and Privacy (SP)}.\hskip 1em plus 0.5em minus 0.4em\relax IEEE, 2024, pp. 862--880.

\bibitem{li2024iris}
Z.~Li, S.~Dutta, and M.~Naik, ``Iris: Llm-assisted static analysis for detecting security vulnerabilities,'' \emph{arXiv preprint arXiv:2405.17238}, 2024.

\bibitem{sun2024llm4vuln}
Y.~Sun, D.~Wu, Y.~Xue, H.~Liu, W.~Ma, L.~Zhang, Y.~Liu, and Y.~Li, ``Llm4vuln: A unified evaluation framework for decoupling and enhancing llms' vulnerability reasoning,'' \emph{arXiv preprint arXiv:2401.16185}, 2024.

\bibitem{deng2024pentestgpt}
G.~Deng, Y.~Liu, V.~Mayoral-Vilches, P.~Liu, Y.~Li, Y.~Xu, T.~Zhang, Y.~Liu, M.~Pinzger, and S.~Rass, ``$\{$PentestGPT$\}$: Evaluating and harnessing large language models for automated penetration testing,'' in \emph{33rd USENIX Security Symposium (USENIX Security 24)}, 2024, pp. 847--864.

\bibitem{li2024enhancing}
H.~Li, Y.~Hao, Y.~Zhai, and Z.~Qian, ``Enhancing static analysis for practical bug detection: An llm-integrated approach,'' \emph{Proceedings of the ACM on Programming Languages}, vol.~8, no. OOPSLA1, pp. 474--499, 2024.

\bibitem{bohme2025software}
M.~B{\"o}hme, E.~Bodden, T.~Bultan, C.~Cadar, Y.~Liu, and G.~Scanniello, ``Software security analysis in 2030 and beyond: A research roadmap,'' \emph{ACM Transactions on Software Engineering and Methodology}, vol.~34, no.~5, pp. 1--26, 2025.

\bibitem{avgustinov2016ql}
P.~Avgustinov, O.~De~Moor, M.~P. Jones, and M.~Sch{\"a}fer, ``Ql: Object-oriented queries on relational data,'' in \emph{30th European Conference on Object-Oriented Programming (ECOOP 2016)}.\hskip 1em plus 0.5em minus 0.4em\relax Schloss Dagstuhl--Leibniz-Zentrum f{\"u}r Informatik, 2016, pp. 2--1.

\bibitem{luo2022tchecker}
C.~Luo, P.~Li, and W.~Meng, ``Tchecker: Precise static inter-procedural analysis for detecting taint-style vulnerabilities in php applications,'' in \emph{Proceedings of the 2022 ACM SIGSAC Conference on Computer and Communications Security}, 2022, pp. 2175--2188.

\bibitem{wang2023conftainter}
T.~Wang, H.~He, X.~Liu, S.~Li, Z.~Jia, Y.~Jiang, Q.~Liao, and W.~Li, ``Conftainter: Static taint analysis for configuration options,'' in \emph{2023 38th IEEE/ACM International Conference on Automated Software Engineering (ASE)}.\hskip 1em plus 0.5em minus 0.4em\relax IEEE, 2023, pp. 1640--1651.

\bibitem{qian2024chatdev}
C.~Qian, W.~Liu, H.~Liu, N.~Chen, Y.~Dang, J.~Li, C.~Yang, W.~Chen, Y.~Su, X.~Cong \emph{et~al.}, ``Chatdev: Communicative agents for software development,'' in \emph{Proceedings of the 62nd Annual Meeting of the Association for Computational Linguistics (Volume 1: Long Papers)}, 2024, pp. 15\,174--15\,186.

\bibitem{kim2022learning}
H.~Kim, M.~Raghothaman, and K.~Heo, ``Learning probabilistic models for static analysis alarms,'' in \emph{Proceedings of the 44th International Conference on Software Engineering}, 2022, pp. 1282--1293.

\bibitem{chen2021boosting}
T.~Chen, K.~Heo, and M.~Raghothaman, ``Boosting static analysis accuracy with instrumented test executions,'' in \emph{Proceedings of the 29th ACM Joint Meeting on European Software Engineering Conference and Symposium on the Foundations of Software Engineering}, 2021, pp. 1154--1165.

\bibitem{synopsys2025coverity}
{Synopsys, Inc.}, ``Coverity scan: Static analysis platform for open source projects,'' \url{https://scan.coverity.com/}, 2025, accessed: 2025-06-26.

\bibitem{yamaguchi2014modeling}
F.~Yamaguchi, N.~Golde, D.~Arp, and K.~Rieck, ``Modeling and discovering vulnerabilities with code property graphs,'' in \emph{2014 IEEE symposium on security and privacy}.\hskip 1em plus 0.5em minus 0.4em\relax IEEE, 2014, pp. 590--604.

\bibitem{livshits2005finding}
V.~B. Livshits and M.~S. Lam, ``Finding security vulnerabilities in java applications with static analysis.'' in \emph{USENIX security symposium}, vol.~14, 2005, pp. 18--18.

\bibitem{tripp2009taj}
O.~Tripp, M.~Pistoia, S.~J. Fink, M.~Sridharan, and O.~Weisman, ``Taj: effective taint analysis of web applications,'' \emph{ACM Sigplan Notices}, vol.~44, no.~6, pp. 87--97, 2009.

\bibitem{arzt2014flowdroid}
S.~Arzt, S.~Rasthofer, C.~Fritz, E.~Bodden, A.~Bartel, J.~Klein, Y.~Le~Traon, D.~Octeau, and P.~McDaniel, ``Flowdroid: Precise context, flow, field, object-sensitive and lifecycle-aware taint analysis for android apps,'' \emph{ACM sigplan notices}, vol.~49, no.~6, pp. 259--269, 2014.

\bibitem{yang2019towards}
J.~Yang, L.~Tan, J.~Peyton, and K.~A. Duer, ``Towards better utilizing static application security testing,'' in \emph{2019 ieee/acm 41st international conference on software engineering: Software engineering in practice (icse-seip)}.\hskip 1em plus 0.5em minus 0.4em\relax IEEE, 2019, pp. 51--60.

\bibitem{christodorescu2007mining}
M.~Christodorescu, S.~Jha, and C.~Kruegel, ``Mining specifications of malicious behavior,'' in \emph{Proceedings of the the 6th joint meeting of the European software engineering conference and the ACM SIGSOFT symposium on The foundations of software engineering}, 2007, pp. 5--14.

\bibitem{chibotaru2019scalable}
V.~Chibotaru, B.~Bichsel, V.~Raychev, and M.~Vechev, ``Scalable taint specification inference with big code,'' in \emph{Proceedings of the 40th ACM SIGPLAN Conference on Programming Language Design and Implementation}, 2019, pp. 760--774.

\bibitem{dong2025fuzz}
Y.~Dong, X.~Meng, N.~Yu, Z.~Li, and S.~Guo, ``Fuzz-testing meets llm-based agents: An automated and efficient framework for jailbreaking text-to-image generation models,'' in \emph{2025 IEEE Symposium on Security and Privacy (SP)}.\hskip 1em plus 0.5em minus 0.4em\relax IEEE, 2025, pp. 373--391.

\bibitem{stafeev2024yurascanner}
A.~Stafeev, T.~Recktenwald, G.~De~Stefano, S.~Khodayari, and G.~Pellegrino, ``Yurascanner: Leveraging llms for task-driven web app scanning,'' 2024.

\bibitem{li2025sv}
Y.~Li, P.~Branco, A.~M. Hoole, M.~Marwah, H.~M. Koduvely, G.-V. Jourdan, and S.~Jou, ``Sv-trusteval-c: Evaluating structure and semantic reasoning in large language models for source code vulnerability analysis,'' in \emph{2025 IEEE Symposium on Security and Privacy (SP)}.\hskip 1em plus 0.5em minus 0.4em\relax IEEE, 2025, pp. 3014--3032.

\bibitem{yang2025context}
Y.~Yang, B.~Xu, X.~Gao, and H.~Sun, ``Context-enhanced vulnerability detection based on large language model,'' \emph{arXiv preprint arXiv:2504.16877}, 2025.

\bibitem{zhou2019devign}
Y.~Zhou, S.~Liu, J.~Siow, X.~Du, and Y.~Liu, ``Devign: Effective vulnerability identification by learning comprehensive program semantics via graph neural networks,'' \emph{Advances in neural information processing systems}, vol.~32, 2019.

\bibitem{lu2024grace}
G.~Lu, X.~Ju, X.~Chen, W.~Pei, and Z.~Cai, ``Grace: Empowering llm-based software vulnerability detection with graph structure and in-context learning,'' \emph{Journal of Systems and Software}, vol. 212, p. 112031, 2024.

\bibitem{sun2024gptscan}
Y.~Sun, D.~Wu, Y.~Xue, H.~Liu, H.~Wang, Z.~Xu, X.~Xie, and Y.~Liu, ``Gptscan: Detecting logic vulnerabilities in smart contracts by combining gpt with program analysis,'' in \emph{Proceedings of the IEEE/ACM 46th International Conference on Software Engineering}, 2024, pp. 1--13.

\bibitem{yang2025knighter}
C.~Yang, Z.~Zhao, Z.~Xie, H.~Li, and L.~Zhang, ``Knighter: Transforming static analysis with llm-synthesized checkers,'' in \emph{Proceedings of the ACM SIGOPS 31st Symposium on Operating Systems Principles}, 2025, pp. 655--669.

\bibitem{bessey2010few}
A.~Bessey, K.~Block, B.~Chelf, A.~Chou, B.~Fulton, S.~Hallem, C.~Henri-Gros, A.~Kamsky, S.~McPeak, and D.~Engler, ``A few billion lines of code later: using static analysis to find bugs in the real world,'' \emph{Communications of the ACM}, vol.~53, no.~2, pp. 66--75, 2010.

\bibitem{codeql_path_queries}
G.~S. Lab, ``Creating path queries,'' \url{https://codeql.github.com/docs/writing-codeql-queries/creating-path-queries/}, accessed: 2025-7-10.

\bibitem{yu2025stateful}
L.~Yu, J.~Lin, and J.~Li, ``Stateful large language model serving with pensieve,'' in \emph{Proceedings of the Twentieth European Conference on Computer Systems}, 2025, pp. 144--158.

\bibitem{agrawal2025llm}
V.~Agrawal and K.~Ahi, ``Llm-driven sast-genius: A hybrid static analysis framework for comprehensive and actionable security,'' \emph{arXiv preprint arXiv:2509.15433}, 2025.

\bibitem{zhong2023scalable}
Z.~Zhong, J.~Liu, D.~Wu, P.~Di, Y.~Sui, A.~X. Liu, and J.~C. Lui, ``Scalable compositional static taint analysis for sensitive data tracing on industrial micro-services,'' in \emph{2023 IEEE/ACM 45th International Conference on Software Engineering: Software Engineering in Practice (ICSE-SEIP)}.\hskip 1em plus 0.5em minus 0.4em\relax IEEE, 2023, pp. 110--121.

\bibitem{bosu2014identifying}
A.~Bosu, J.~C. Carver, M.~Hafiz, P.~Hilley, and D.~Janni, ``Identifying the characteristics of vulnerable code changes: An empirical study,'' in \emph{Proceedings of the 22nd ACM SIGSOFT international symposium on foundations of software engineering}, 2014, pp. 257--268.

\bibitem{kharkar2022learning}
A.~Kharkar, R.~Z. Moghaddam, M.~Jin, X.~Liu, X.~Shi, C.~Clement, and N.~Sundaresan, ``Learning to reduce false positives in analytic bug detectors,'' in \emph{Proceedings of the 44th International Conference on Software Engineering}, 2022, pp. 1307--1316.

\bibitem{li2018vuldeepecker}
Z.~Li, D.~Zou, S.~Xu, X.~Ou, H.~Jin, S.~Wang, Z.~Deng, and Y.~Zhong, ``Vuldeepecker: A deep learning-based system for vulnerability detection,'' \emph{arXiv preprint arXiv:1801.01681}, 2018.

\bibitem{li2021sysevr}
Z.~Li, D.~Zou, S.~Xu, H.~Jin, Y.~Zhu, and Z.~Chen, ``Sysevr: A framework for using deep learning to detect software vulnerabilities,'' \emph{IEEE Transactions on Dependable and Secure Computing}, vol.~19, no.~4, pp. 2244--2258, 2021.

\bibitem{cao2025recurring}
Y.~Cao, S.~Wu, R.~Wang, B.~Chen, Y.~Huang, C.~Lu, Z.~Zhou, and X.~Peng, ``Recurring vulnerability detection: How far are we?'' \emph{Proceedings of the ACM on Software Engineering}, vol.~2, no. ISSTA, pp. 573--595, 2025.

\bibitem{bohme2017directed}
M.~B{\"o}hme, V.-T. Pham, M.-D. Nguyen, and A.~Roychoudhury, ``Directed greybox fuzzing,'' in \emph{Proceedings of the 2017 ACM SIGSAC conference on computer and communications security}, 2017, pp. 2329--2344.

\bibitem{iris_sast_cwe_bench_java}
IRIS-SAST, ``Cwe-bench-java: A manually vetted dataset for security vulnerability detection in java projects,'' \url{https://github.com/iris-sast/cwe-bench-java}, 2024, gitHub repository.

\bibitem{openai_gpt4_1_blog2025}
OpenAI, ``Introducing gpt-4.1 in the api,'' \url{https://openai.com/index/gpt-4-1/}, Apr. 2025, accessed: 2025-6-13.

\bibitem{deepseek_r1_0528_2025}
D.~AI, ``Deepseek-r1-0528: Model card and release notes,'' \url{https://huggingface.co/deepseek-ai/DeepSeek-R1-0528}, 2025, accessed: 2025-6-16.

\bibitem{z_ai_glm4.5_blog}
Z.~Team, ``Glm-4.5: Reasoning, coding, and agentic abilities,'' \url{https://z.ai/blog/glm-4.5}, July 2025.

\bibitem{qwen_ai_qwen3_blog}
Q.~Team, ``Qwen3: Think deeper, act faster,'' \url{https://qwen.ai/blog?id=qwen3}, April 2025.

\bibitem{10646663}
S.~Ullah, M.~Han, S.~Pujar, H.~Pearce, A.~Coskun, and G.~Stringhini, ``Llms cannot reliably identify and reason about security vulnerabilities (yet?): A comprehensive evaluation, framework, and benchmarks,'' in \emph{2024 IEEE Symposium on Security and Privacy (SP)}, 2024, pp. 862--880.

\bibitem{jovanovic2006pixy}
N.~Jovanovic, C.~Kruegel, and E.~Kirda, ``Pixy: A static analysis tool for detecting web application vulnerabilities,'' in \emph{2006 IEEE Symposium on Security and Privacy (S\&P'06)}.\hskip 1em plus 0.5em minus 0.4em\relax IEEE, 2006, pp. 6--pp.

\bibitem{liu2023trustworthy}
Y.~Liu, Y.~Yao, J.-F. Ton, X.~Zhang, R.~Guo, H.~Cheng, Y.~Klochkov, M.~F. Taufiq, and H.~Li, ``Trustworthy llms: a survey and guideline for evaluating large language models' alignment,'' \emph{arXiv preprint arXiv:2308.05374}, 2023.

\bibitem{tonmoy2024comprehensive}
S.~Tonmoy, S.~Zaman, V.~Jain, A.~Rani, V.~Rawte, A.~Chadha, and A.~Das, ``A comprehensive survey of hallucination mitigation techniques in large language models,'' \emph{arXiv preprint arXiv:2401.01313}, vol.~6, 2024.

\bibitem{pan-etal-2023-logic}
\BIBentryALTinterwordspacing
L.~Pan, A.~Albalak, X.~Wang, and W.~Wang, ``Logic-{LM}: Empowering large language models with symbolic solvers for faithful logical reasoning,'' in \emph{Findings of the Association for Computational Linguistics: EMNLP 2023}, H.~Bouamor, J.~Pino, and K.~Bali, Eds.\hskip 1em plus 0.5em minus 0.4em\relax Singapore: Association for Computational Linguistics, Dec. 2023, pp. 3806--3824. [Online]. Available: \url{https://aclanthology.org/2023.findings-emnlp.248/}
\BIBentrySTDinterwordspacing

\bibitem{morishita2024enhancing}
T.~Morishita, G.~Morio, A.~Yamaguchi, and Y.~Sogawa, ``Enhancing reasoning capabilities of llms via principled synthetic logic corpus,'' \emph{Advances in Neural Information Processing Systems}, vol.~37, pp. 73\,572--73\,604, 2024.

\bibitem{lin2025zebralogic}
B.~Y. Lin, R.~L. Bras, K.~Richardson, A.~Sabharwal, R.~Poovendran, P.~Clark, and Y.~Choi, ``Zebralogic: On the scaling limits of llms for logical reasoning,'' \emph{arXiv preprint arXiv:2502.01100}, 2025.

\bibitem{9700318}
U.~Yüksel and H.~Sözer, ``Dynamic filtering and prioritization of static code analysis alerts,'' in \emph{2021 IEEE International Symposium on Software Reliability Engineering Workshops (ISSREW)}, 2021, pp. 294--295.

\bibitem{gelman2023escalatedquicklymlframework}
\BIBentryALTinterwordspacing
B.~Gelman, S.~Taoufiq, T.~Vörös, and K.~Berlin, ``That escalated quickly: An ml framework for alert prioritization,'' 2023. [Online]. Available: \url{https://arxiv.org/abs/2302.06648}
\BIBentrySTDinterwordspacing

\bibitem{279967}
\BIBentryALTinterwordspacing
W.~Li, J.~Ming, X.~Luo, and H.~Cai, ``{PolyCruise}: A {Cross-Language} dynamic information flow analysis,'' in \emph{31st USENIX Security Symposium (USENIX Security 22)}.\hskip 1em plus 0.5em minus 0.4em\relax Boston, MA: USENIX Association, Aug. 2022, pp. 2513--2530. [Online]. Available: \url{https://www.usenix.org/conference/usenixsecurity22/presentation/li-wen}
\BIBentrySTDinterwordspacing

\bibitem{10.1145/3468264.3468538}
\BIBentryALTinterwordspacing
G.~Mathew and K.~T. Stolee, ``Cross-language code search using static and dynamic analyses,'' in \emph{Proceedings of the 29th ACM Joint Meeting on European Software Engineering Conference and Symposium on the Foundations of Software Engineering}, ser. ESEC/FSE 2021.\hskip 1em plus 0.5em minus 0.4em\relax New York, NY, USA: Association for Computing Machinery, 2021, p. 205–217. [Online]. Available: \url{https://doi.org/10.1145/3468264.3468538}
\BIBentrySTDinterwordspacing

\bibitem{zhang2025abusability}
S.~Zhang, P.~Chung, J.~Vervelde, N.~Korapati, R.~Chatterjee, and K.~Fawaz, ``Abusability of automation apps in intimate partner violence,'' in \emph{34th USENIX Security Symposium (USENIX Security 25)}, 2025, pp. 41--60.

\bibitem{hasegawa2024weird}
A.~A. Hasegawa, D.~Inoue, and M.~Akiyama, ``How $\{$WEIRD$\}$ is usable privacy and security research?'' in \emph{33rd USENIX Security Symposium (USENIX Security 24)}, 2024, pp. 3241--3258.

\bibitem{first2020cvd}
{FIRST.org}, ``Guidelines and practices for multi-party vulnerability coordination and disclosure (version 1.1),'' \url{https://www.first.org/global/sigs/vulnerability-coordination/multiparty/FIRST-Multiparty-Vulnerability-Coordination-v1.1.pdf}, 2020.

\end{thebibliography}

\end{document}